%Paper: hep-ph/9211286
%From: My Account <me@cryptons.tamu.edu>
%Date: Fri, 20 Nov 92 10:50:07 -0800

% This paper uses macropackages harvmac.tex and tables.tex.
% Type 'b' if confused.
% The 14 postscript figures are sent separately in uuencoded form as required
% by the new command 'figures'. In a unix system save the figures file to
% say uufigs and remove the mail header. Then do: chmod +x uufigs
% and execute it (ie, just type uufigs). The file will then unpack itself into
%%% 14 ps files named lnw*.ps. Finally do lpr lnw*.ps to get the hard copies.
%
\input harvmac

\def\Titlehh#1#2{\nopagenumbers\abstractfont\hsize=\hstitle\rightline{#1}%
\vskip .2in\centerline{\titlefont #2}\abstractfont\vskip .2in\pageno=0}
\def\CTPa{\it Center for Theoretical Physics, Department of Physics,
      Texas A\&M University}
\def\CTPb{\it College Station, TX 77843-4242, USA}
\def\HARCa{\it Astroparticle Physics Group,
Houston Advanced Research Center (HARC)}
\def\HARCb{\it The Woodlands, TX 77381, USA}

\def\ie{\hbox{\it i.e.}}     
\def\eg{\hbox{\it e.g.}}     

\def\nextline{\unskip\nobreak\hfill\break}

\catcode`\@=11 % This allows us to modify PLAIN macros.

\def\lsim{\mathrel{\mathpalette\@versim<}}
\def\gsim{\mathrel{\mathpalette\@versim>}}
\def\@versim#1#2{\vcenter{\offinterlineskip
    \ialign{$\m@th#1\hfil##\hfil$\crcr#2\crcr\sim\crcr } }}
\def\boxit#1{\vbox{\hrule\hbox{\vrule\kern3pt
      \vbox{\kern3pt#1\kern3pt}\kern3pt\vrule}\hrule}}

\def\r#1{$\bf#1$}
\def\rb#1{$\bf\overline{#1}$}

\def\t1{{\tilde 1}}

\def\JL{J. L. Lopez}
\def\DVN{D. V. Nanopoulos}

\def\GeV{\,{\rm GeV}}
\def\TeV{\,{\rm TeV}}

\def\pb{\,{\rm pb}}
\def\ipb{\,{\rm pb^{-1}}}

\def\wt{\widetilde}

\def\NPB#1#2#3{Nucl. Phys. B {\bf#1} (19#2) #3}
\def\PLB#1#2#3{Phys. Lett. B {\bf#1} (19#2) #3}

\def\PRL#1#2#3{Phys. Rev. Lett. {\bf#1} (19#2) #3}
\def\PRT#1#2#3{Phys. Rep. {\bf#1} (19#2) #3}
\def\MODA#1#2#3{Mod. Phys. Lett. A {\bf#1} (19#2) #3}

\def\TAMU#1{Texas A \& M University preprint CTP-TAMU-#1}

\nref\LN{For a review see A. Lahanas and \DVN, \PRT{145}{87}{1}.}
\nref\AN{R. Arnowitt and P. Nath, \PRL{69}{92}{725}; P. Nath and
R. Arnowitt, \PLB{287}{92}{89} and \PLB{289}{92}{368}.}
\nref\LNPZ{\JL, \DVN, H. Pois, and A. Zichichi, \TAMU{72/92},
CERN-TH.6716/92, and CERN-PPE/92-189.}
\nref\LNZa{\JL, \DVN, and A. Zichichi, \PLB{291}{92}{255}.}
\nref\LNP{\JL, \DVN, and H. Pois, \TAMU{61/92} and CERN-TH.6628/92.}
\nref\LNZb{\JL, \DVN, and A. Zichichi, \TAMU{68/92}, CERN-TH.6667/92, and
CERN-PPE/92-188.}
\nref\ANch{P. Nath and R. Arnowitt, \MODA{2}{87}{331}.}
\nref\Barbieri{R. Barbieri, F. Caravaglios, M. Frigeni, and M. Mangano,
\NPB{367}{91}{28}.}
\nref\BT{H. Baer and X. Tata, Florida preprint FSU-HEP-920907.}
\nref\aspects{For a detailed description of this general procedure see \eg,
S. Kelley, \JL, \DVN, H. Pois, and K. Yuan, \TAMU{16/92} and CERN-TH.6498/92.}
\nref\JHW{For a review see \JL\ and \DVN, in Proceedings of
the 15th Johns Hopkins Workshop on Current Problems in Particle Theory, August
1991, p. 277; ed. by G. Domokos and S. Kovesi-Domokos.}
\nref\search{\JL, \DVN, and K. Yuan, \TAMU{11/92}.}
\nref\ECNsun{J. Ellis, C. Kounnas, and \DVN, \NPB{247}{84}{373}.}
\nref\Lahanas{J. Ellis, A. Lahanas, \DVN, and K. Tamvakis, \PLB{134}{84}{429}.}
\nref\MT{J. G. Morfin and W. K. Tung, Z. Phys. {\bf C52} (1991) 13.}
\nref\JTW{J. White, private communication.}

\nfig\I{The cross section for $p\bar p\to\chi^\pm_1\chi^0_2$ versus the
chargino mass at $\sqrt{s}=1.8\TeV$ for all points in the allowed parameter
space of (a) the minimal $SU(5)$ supergravity model and (b) the no-scale
flipped $SU(5)$ supergravity model.}
\nfig\II{The branching ratio for $\chi^\pm_1\to\chi^0_1\nu_e e^\pm,\chi^0_1
\nu_\mu\mu^\pm$ in the no-scale flipped $SU(5)$ supergravity model for (a)
$m_t=100\GeV$, (b) $m_t=130\GeV$, and (c) $m_t=160\GeV$. Note that the
branching ratio is bounded between $2/9$ (when the $W$-exchange diagrams
dominate) and $2/3$ (when the slepton exchange diagrams dominate).}
\nfig\III{The branching ratio for
$\chi^0_2\to\chi^0_1e^+e^-,\chi^0_1\mu^+\mu^-$ in the minimal $SU(5)$
supergravity model. The points on the horizontal axis occur when the spoiler
mode is open. The accumulation of points around $6.6\%$ occurs when the
$Z$-exchange diagrams dominate.}
\nfig\IV{The branching ratio for $\chi^0_2\to\chi^0_1e^+e^-,\chi^0_1\mu^+\mu^-$
in the no-scale flipped $SU(5)$ supergravity model for (a) $m_t=100\GeV$, (b)
$m_t=130\GeV$, and (c) $m_t=160\GeV$. The accumulation of points near the
horizontal axis correspond to a new spoiler mode
$\chi^0_2\to\chi^0_1\tilde\nu_l$. The maximum values are reached when the
`anti-spoiler mode' $\chi^0_2\to\chi^0_1 l^\pm\tilde l^\mp_{L,R}$ opens up.}
\nfig\V{The number of trilepton events per $100\ipb$ for the minimal $SU(5)$
supergravity model (excluding points for which the spoiler mode is open).
Note that with $200\ipb$ and $60\%$ efficiency it should be possible to probe
up to $m_{\chi^\pm_1}=90-95\GeV$.}
\nfig\VI{The number of trilepton events per $100\ipb$ for the no-scale flipped
$SU(5)$ supergravity model (excluding points for which the spoiler modes are
open) for (a) $m_t=100\GeV$, (b) $m_t=130\GeV$, and (c) $m_t=160\GeV$.
Note that with $200\ipb$ and $60\%$ efficiency it should be possible to probe
up to $m_{\chi^\pm_1}\approx200\GeV$.}
\nfig\VII{The number of trilepton events per $100\ipb$ for the strict case of
the no-scale flipped $SU(5)$ supergravity model for the indicated values of
the top-quark mass in GeV. The spoiler modes are only
open for $m_{\chi^\pm_1}\lsim75\GeV$. Note that with $200\ipb$ and $60\%$
efficiency it should be possible to {\it fully} probe
$75\GeV\lsim m_{\chi^\pm_1}\lsim200\GeV$.}

\Titlehh{\vbox{\baselineskip12pt
\hbox{CERN/LAA/92--023}
\hbox{CTP--TAMU--76/92}\hbox{ACT--23/92}}}
{\vbox{\centerline{Supersymmetry Tests at Fermilab: A Proposal}}}
\centerline{JORGE~L.~LOPEZ$^{(a)(b)}$, D.~V.~NANOPOULOS$^{(a)(b)}$,
XU WANG$^{(a)(b)}$, and A. ZICHICHI$^{(c)}$}
\smallskip
\centerline{$^{(a)}$\CTPa}
\centerline{\CTPb}
\centerline{$^{(b)}$\HARCa}
\centerline{\HARCb}
\centerline{$^{(c)}${\it CERN, Geneva, Switzerland}}
\vskip .2in
\centerline{ABSTRACT}
We compute the number of trilepton events to be expected at Fermilab
as a result of the reaction $p\bar p\to\chi^\pm_1\chi^0_2X$, where $\chi^\pm_1$
is the lightest chargino and $\chi^0_2$ is the next-to-lightest neutralino.
This signal is expected to have very little background and is the best
prospect for supersymmetry detection at Fermilab if the gluino and squarks
are beyond reach. We evaluate our expressions for all points in the allowed
parameter space of two basic supergravity models: (i) the minimal $SU(5)$
supergravity model including the severe constraints from proton decay and a not
too young Universe, and (ii) a recently proposed no-scale flipped $SU(5)$
supergravity model. We study the plausible experimental scenaria and
conclude that a large portion of the parameter spaces of these models could
be explored with $100\ipb$ of integrated luminosity. In the minimal $SU(5)$
supergravity model chargino masses up to $90-95\GeV$ could be probed.
In the no-scale flipped model it should be possible to probe some regions of
parameter space with $m_{\chi^\pm_1}\lsim200\GeV$, therefore possibly doubling
the reach of LEPII for chargino and neutralino (since $m_{\chi^0_2}\approx
m_{\chi^\pm_1}$) masses. In both models such probes would indirectly explore
gluino masses much beyond the reach of Fermilab.
\bigskip
\bigskip
\bigskip
\bigskip
{\vbox{\baselineskip12pt
\hbox{CERN/LAA/92--023}
\hbox{CTP--TAMU--76/92}
\hbox{ACT--23/92}}}
\Date{November, 1992}

\newsec{Introduction}
It is becoming ever more apparent that supersymmetry is ``the way to go" beyond
the Standard Model. Besides its numerous theoretical niceties -- such as its
role in solving the hierarchy problem, in explaining the lightness of the
Higgs boson, in the unification of the gauge couplings, in the unification
with gravity, and in superstrings -- supersymmetry entails a rather predictive
and experimentally appealing framework. On the most general grounds however,
all one can say is that we expect the set of superpartners of the ordinary
particles to appear somewhere below $\sim1\TeV$. Moreover, the number of
parameters needed to describe the new particles is rather large (at least
twenty-one), and therefore a full exploration of the parameter space of a
generic low-energy supersymmetric model is impractical.

On the other hand,
things become much simpler if one studies specific models which embody a
minimal set of well motivated theoretical assumptions, including spontaneously
broken supergravity with universal soft-supersymmetry breaking, and
radiative electroweak symmetry breaking \LN. In fact, the dimension of the
parameter space of these models is quite minimal: three soft-supersymmetry
breaking masses ($m_{1/2},m_0,A$), the ratio of Higgs vacuum expectation
values ($\tan\beta$), and the top-quark mass ($m_t$). Among these models
there are two which stand out because of their high predictive power:
(i) the minimal $SU(5)$ supergravity model including the severe constraints
from proton decay \refs{\AN,\LNPZ} and a not too young Universe
\refs{\LNZa,\LNP}, and (ii) a recently proposed no-scale flipped $SU(5)$
supergravity model \LNZb. In fact, these two models can be viewed as prototypes
of (i) traditional supergravity GUT models and (ii) string-inspired
supergravity models, respectively.

In this paper we begin our study of supersymmetric signals from these two
models at the Fermilab Tevatron collider by considering the trilepton signal
which occurs in the decay products of the reaction
$p\bar p\to \chi^\pm_1\chi^0_2X$, where $\chi^\pm_1$ is the lightest chargino
and $\chi^0_2$ is the next-to-lightest neutralino. This hadronically quiet
signal has been shown to have very little background \refs{\ANch,\Barbieri,\BT}
and is expected to be the best one for exploring the neutralino-chargino sector
of supersymmetric models at Fermilab \BT. In fact, even if the squark and
gluino masses are well
beyond the reach of Fermilab, the two models under consideration still predict
trilepton signals that can be directly observable. Conversely, the potential
exclusion of certain regions of the neutralino-chargino parameter space would
entail indirect exploration of a large range of squark and gluino masses.

We first compare the features of both models and their predicted
supersymmetric spectra. We then compute in succession the
$p\bar p\to \chi^\pm_1\chi^0_2X$ cross section, the branching ratio of
$\chi^\pm_1$ to one charged lepton, and the branching ratio of $\chi^0_2$ to
two charged leptons. We find that the branching ratios in the no-scale
flipped $SU(5)$ supergravity model depend crucially on the slepton mass
spectrum since it can be relatively light, and therefore differ significantly
from the standard results which usually assume heavy sleptons. We conclude that
with $100\ipb$ of integrated luminosity and optimal detection efficiencies, it
should be possible to explore a significant portion of the allowed parameter
space of both models.

\newsec{The Models}
Both models under study have the same light matter content as the minimal
supersymmetric extension of the Standard Model. That is, besides the ordinary
particles, we have: (i) twelve squarks $\tilde u_{L,R},\tilde d_{L,R},
\tilde c_{L,R}, \tilde s_{L,R}, \tilde b_{1,2}, \tilde t_{1,2}$, (ii) nine
sleptons $\tilde e_{L,R},\tilde\mu_{L,R},\tilde
\tau_{1,2},\tilde\nu_{e,\mu,\tau}$, (iii) two charginos $\chi^\pm_{1,2}$ and
four neutralinos $\chi^0_{1,2,3,4}$, and (iv) two CP-even neutral Higgs bosons
$h,H$, one CP-odd neutral Higgs $A$, and one charged Higgs $H^\pm$. The masses
of all these particles depend on a set of parameters which can be significantly
reduced by imposing universal soft-supersymmetry breaking at the unification
scale and then radiative electroweak symmetry breaking \aspects. Once this is
done the masses depend on only five parameters: $m_{1/2},\xi_0\equiv
m_0/m_{1/2},\xi_A\equiv A/m_{1/2},\tan\beta,m_t$. For
reference, the squark and slepton masses can be approximated by
$m^2_i\approx({1\over3}m_{\tilde g})^2(c_i+\xi^2_0)+d_i\cos2\beta M^2_Z$, where
$c_i$ and $d_i$ are calculable coefficients with
$0<c_{\tilde l}\ll c_{\tilde q}\lsim6$, and $|d_i|\lsim1$. The fermion mass
contribution to the respective sfermion mass is neglected except for $\tilde
b_{1,2},\tilde t_{1,2},\tilde\tau_{1,2}$. The first and second generation
squarks are nearly
degenerate in mass, while the sbottom and stop mass eigenstates are split
relative to the average squark mass. For the sleptons, near degeneracy only
occurs for $\xi_0\gg1$. The chargino and
neutralino masses depend on $\tan\beta, m_{\tilde g}$, and the Higgs mixing
parameter $\mu$, whose magnitude is calculable from the radiative breaking
constraints, but its sign remains undetermined. The Higgs boson masses receive
a tree-level contribution which depends on $\tan\beta$ and $m_A$, and a
one-loop correction which depends most importantly on $m_t$ and the squark
mass.  For any given set $(m_{1/2},\xi_0,\xi_A,\tan\beta,m_t)$ one can compute
all particle masses and couplings and reject sets which violate present
experimental bounds on
$m_{\tilde g},m_{\tilde q},m_h,m_A,m_{\chi^\pm_1},\Gamma^{inv}_Z$, etc.
Only those sets which satisfy all present phenomenological constraints (as
described in detail in Ref. \aspects) are kept for further analysis. All the
above remarks apply to the two models under consideration. We now turn to the
differences between them. For reference, in Table I we collect the gist of the
following discussion.
\subsec{The minimal $SU(5)$ supergravity model}
This model is based on the gauge group $SU(5)$ and its minimal matter content
implies that it unifies at $M_U\sim10^{16}\GeV$. In the context of string
theory, this model is in principle derivable but in practice this is hard to
accomplish and no known examples exist. Symmetry breaking down to the Standard
Model occurs through the vev of the \r{24} of Higgs. The light Higgs doublets
come together with colored Higgs triplets ($H_3$) in $SU(5)$ pentaplets. The
$H_3$ fields mediate dangerous dimension-five proton decay operators and must
be heavy ($M_{H_3}\gsim M_U$). This needed ``doublet-triplet splitting" is not
explained in a simple way in this model and requires either an ad-hoc choice
of parameters or the introduction of additional large GUT-mass representations.
Requiring that $M_{H_3}<3M_U$ so that the high-energy theory remains
perturbative, and imposing the naturalness condition $m_{\tilde g},m_{\tilde q}
<1\TeV$, it has been shown that the five-dimensional parameter space of this
model is severely constrained by the experimental lower bound on the proton
lifetime \AN. Adding on the requirement of a not too young Universe --
$\Omega_\chi h^2_0\le1$, where $\Omega_\chi$ is the present abundance of the
lightest neutralino and $0.5\le h_0\le1$ is the scaled Hubble parameter --
constrains the model significantly more \refs{\LNZa,\LNP}. Yet stricter
constraints follow from the recalculation of the unification mass to a higher
level of precision \LNPZ. The finally allowed set of points
in parameter space is highly restricted and entails the following constraints:
(i) parameter space variables
\eqn\I{1\lsim\tan\beta\lsim3.5, \quad m_t\lsim180\GeV,\quad \xi_0\gsim6,}
(ii) gluino, squark, and slepton masses
\eqn\II{m_{\tilde g}\lsim500\GeV,\quad
 m_{\tilde q}>m_{\tilde l}>2m_{\tilde g},}
\eqna\III
(iii) chargino and neutralino masses\foot{The $\sim$ signs in Eqs. \III{} and
(2.7) indicate that these relations are only qualitative, although the
majority of points in the allowed parameter space follow them closely.}
$$\eqalignno{&2m_{\chi^0_1}\sim m_{\chi^0_2}\sim m_{\chi^\pm_1}\sim
0.3 m_{\tilde g}\lsim150\GeV,&\III a\cr
&m_{\chi^0_3}\sim m_{\chi^0_4}\sim m_{\chi^\pm_2}\sim|\mu|,&\III b\cr}$$
(iv) the one-loop corrected lightest Higgs boson mass
\eqn\IV{m_h\lsim100\GeV.}

\subsec{The no-scale flipped $SU(5)$ supergravity model {\rm\LNZb}}
This recently proposed model \LNZb\ is based on the gauge group $SU(5)\times
U(1)$ and has additional intermediate scale matter particles that delay
unification until $M_U\sim10^{18}\GeV$, as expected to occur in string-derived
models. The minimal choice of the extra particles is a pair of vector-like
quark doublets $Q,\bar Q$ with $m_Q\sim10^{12}\GeV$ and a pair of vector-like
charge $-1/3$ quark singlets $D,\bar D$ with $m_D\sim10^6\GeV$. There exist
several string models based on this gauge group \JHW\ and at least one with
the additional matter particles \search. Symmetry breaking down to the
Standard Model gauge group occurs through vevs of \r{10},\rb{10}
representations along flat directions of the scalar potential, and thus it is
tied to the onset of supersymmetry breaking. The doublet-triplet splitting
of the Higgs pentaplets alluded to in Sec. 2.1 is realized naturally in this
model through gauge symmetry allowed couplings which occur in all known
examples. This mechanism also ensures that the potentially dangerous
dimension-five proton decay operators are highly suppressed and innocuous.
The no-scale supergravity component of the model implies that $m_0=A=0$
\ECNsun\ and therefore the model depends only on three parameters: $m_{1/2},
\tan\beta,m_t$. Furthermore, consistency of the no-scale model  requires
$m_{1/2}\lsim1\TeV$ \Lahanas\ which explains the naturalness requirement which
otherwise would need to be imposed by hand. The relic abundance of the lightest
neutralino is found to be $\Omega_\chi h^2_0\lsim0.25$, which is well within
cosmological requirements and large enough to explain the dark matter problem.

This model also entails constraints on its parameters and correlations among
the various particle masses:\nextline
(i) parameter space variables
\eqn\If{2\lsim\tan\beta\lsim32,\quad
m_t\lsim190\GeV,\quad\xi_0=0,\quad\xi_A=0,}
(ii) gluino, squark, and slepton masses
\eqna\IIf
$$\eqalignno{&m_{\tilde g}\lsim1\TeV,\quad m_{\tilde q}\approx m_{\tilde
g},&\IIf a\cr
&m_{\tilde l_L}\approx m_{\tilde\nu}\approx0.3m_{\tilde g}\lsim300\GeV,&\IIf
b\cr
&m_{\tilde l_R}\approx0.18 m_{\tilde g}\lsim200\GeV,&\IIf c\cr}$$
(iii) chargino and neutralino masses
\eqna\IIIf
$$\eqalignno{&2m_{\chi^0_1}\sim m_{\chi^0_2}\approx m_{\chi^\pm_1}\sim
0.3 m_{\tilde g}\lsim285\GeV,&\IIIf a\cr
&m_{\chi^0_3}\sim m_{\chi^0_4}\sim m_{\chi^\pm_2}\sim|\mu|,&\IIIf b\cr}$$
(iv) the one-loop corrected lightest Higgs boson mass
\eqn\IVf{m_h\lsim135\GeV.}

In addition, a strict version of the no-scale scenario allows $\tan\beta$ to
be determined as a function of $m_{\tilde g}$ and $m_t$. This special case
of ``no-scale" has two very interesting consequences: (i) determination of the
sign of $\mu$ and (ii) determination of whether $m_h$ is above or below
$100\GeV$. One finds that $\mu>0$ and $m_h\lsim100\GeV$ if $m_t\lsim135\GeV$,
whereas $\mu<0$ and $m_h\gsim100\GeV$ if $m_t\gsim140\GeV$.

\newsec{The trilepton signal}
The set of diagrams that needs to be calculated is the same for both models,
only the input masses and couplings differ, and so do the resulting signals.
Two diagrams contribute to $p\bar p\to\chi^+_1\chi^0_2X$: (i) $s$-channel
virtual $W$ exchange $u\bar d\to W^* \to \chi^+_1\chi^0_2$, and (ii)
$t$-channel squark exchange. The second diagram has been neglected since
$m_{\tilde q}\gsim200\GeV$ in the no-scale flipped $SU(5)$ model and
$m_{\tilde q}\gsim600\GeV$ in the minimal $SU(5)$ model, and the $W\chi^\pm_1
\chi^0_2$ coupling only vanishes if $\chi^0_2$ is a pure bino which does not
occur in practice. In Figs. 1a and 1b we show the cross section (summed over
$\chi^+_1\chi^0_2$ and $\chi^-_1\chi^0_2$) for $\sqrt{s}=1.8\TeV$, computed
using the parton distribution functions of Ref. \MT. This set of parton
distribution functions is given in a convenient, compact, analytical form and
describes well the small-$x$ behavior.
For the points in parameter space in common with Ref. \BT, we have checked
that our numerical results agree well with theirs.\foot{Note that our sign
convention for $\mu$ is opposite to that used in Ref. \BT.}
The scatter plots include all allowed points in parameter space as obtained
in Ref. \LNPZ\ for the minimal $SU(5)$ model and in Ref. \LNZb\ for the
no-scale flipped model.\foot{For the no-scale flipped model, in Ref. \LNZb\
$\alpha_3(M_Z)=0.118$ was used. For the minimal $SU(5)$ model, in Ref. \LNPZ\
$\alpha_3(M_Z)=0.126$ was instead chosen  in order to maximize
the proton lifetime and therefore the size of the allowed parameter space.}
In the former case $m_t$ takes values throughout the
interval $100-160\GeV$, whereas in the latter case only  the reference values
$m_t=100,130,160\GeV$ are shown (since there are many more allowed points in
parameter space). From the figure one can see that in both models
\eqn\TI{\sigma(p\bar p\to\chi^\pm_1\chi^0_2X)\gsim1\pb\quad{\rm for}\quad
m_{\chi^\pm_1}\lsim100\GeV.}
One can also show that for $m_{\chi^\pm_1}<100\GeV$ the {\it maximum}
indirectly explorable gluino masses are given by: (i) $320\,(460)\GeV$ (for
$\mu>0\,(\mu<0)$) in the minimal $SU(5)$ case, and (ii) $490\,(515)\GeV$ in
the no-scale flipped case.

The $\chi^\pm_1$ decay channels which are kinematically open in these models
are: (i) $\chi^\pm_1\to \chi^0_1 \nu_l l^\pm$, which proceeds through virtual
$W$, $\tilde l_L$, and $\tilde\nu_l$ exchanges, and (ii) $\chi^\pm_1\to
\chi^0_1 q\bar q'$, through virtual $W$ and $\tilde q_L$ exchanges. Channels
with $\chi^0_2$ in the final state are either closed or phase space suppressed
(since $m_{\chi^0_2}\approx m_{\chi^\pm_1}$) and have been neglected. If the
sleptons and squarks are heavy enough, the $W$-exchange diagrams dominate and
the branching ratio into $e^\pm$ and $\mu^\pm$ should be close to
$(1+1)/(1+1+1+3+3)=2/9\approx22\%$, independently of the model
parameters.\foot{This assumes that the $\chi^+_1\to\chi^0_1 t\bar b$ channel
is closed (\ie, $m_{\chi^\pm_1}<m_{\chi^0_1}+m_t+m_b$) which is certainly true
for $m_{\chi^\pm_1}<100\GeV$. Using the approximate relations in Eqs. \III{a}
and \IIIf{a} this channel will not be open until $m_{\chi^\pm_1}\gsim2(m_t+m_b)
\gsim190\GeV$.} This is precisely the result we obtained for the minimal
$SU(5)$ supergravity model,
\eqn\TII{{\rm BR}
(\chi^\pm_1\to \chi^0_1\nu_e e^\pm,\chi^0_1\nu_\mu \mu^\pm)_{minimal}
\approx0.222\pm0.008.}

The results for the no-scale flipped $SU(5)$ model are shown in Fig. 2 for
(a) $m_t=100\GeV$, (b) $m_t=130\GeV$, and (c) $m_t=160\GeV$. We find
\eqn\TIII{2/9\lsim{\rm BR}(\chi^\pm_1\to \chi^0_1\nu_e e^\pm,\chi^0_1\nu_\mu
\mu^\pm)_{flipped}\lsim2/3.}
The upper bound is model-independent and occurs when the slepton exchange
diagrams dominate (\eg, when the sleptons go on-shell), in particular the
sneutrino one since $m_{\tilde\nu}<m_{\tilde e_L}$ for low chargino masses.
As $m_{1/2}$ grows, all masses grow and the sleptons become heavier than the
$W$ boson. Even in this case can one still have slepton dominance since the
$W\chi^\pm_1\chi^0_1$ coupling goes to zero in some ``pure" limits,
\ie, when $\chi^\pm_1\to \wt W^\pm$ and $\chi^0_1\to\wt B$. In fact, Fig. 2
shows that as $m_t$ grows, the branching ratios tend to approach the $2/3$
value. This is understandable since $|\mu|$ grows with $m_t$ (for fixed
$m_{\tilde g}$) and the ``pure" limits are then approached.

The $\chi^0_2$ decay channels which are possibly kinematically open are
several:
(i) $\chi^0_2\to\chi^0_1 l^+l^-$, through virtual $Z, \tilde l_L,\tilde l_R$
exchanges, (ii) $\chi^0_2\to\chi^0_1\nu_l\bar\nu_l$, through $Z,\tilde\nu_l$,
(iii) $\chi^0_2\to\chi^0_1 q\bar q$, through $Z,\tilde q_L,\tilde q_R$, and
(iv) the `spoiler mode' $\chi^0_2\to\chi^0_1 h$. In analogy with the chargino
decay, if the sleptons and squarks are heavy enough, and the
$Z\chi^0_1\chi^0_2$ coupling is not too small, then the $Z$-exchange diagrams
dominate and ${\rm BR}(\chi^0_2\to\chi^0_1 e^+e^-,\chi^0_1\mu^+\mu^-)
\approx{\rm BR}(Z\to e^+e^-,\mu^+\mu^-)\approx6.6\%$. In the minimal $SU(5)$
supergravity model this situation occurs quite often, see Fig. 3. In fact,
from Fig. 3 (bottom row) in Ref. \LNP\ one can see that the $\chi^0_1$
composition is away from  a pure bino state (which would drive the
$Z\chi^0_1\chi^0_2$ coupling to zero) for most of the parameter space.
Exceptions to this situation occur for $\mu>0$ when $m_{\tilde g}$ is low
(and thus so is $m_{\chi^\pm_1}$), while for $\mu<0$ when $m_{\tilde g}\sim
300\GeV\Rightarrow m_{\chi^\pm_1}\sim90\GeV$.  Fig. 3 here shows that in these
cases the branching ratio is higher, since the $Z$-exchange
channels are highly suppressed and the slepton and squark exchanges play a
role. The largest value the branching ratio could take is $\approx2/3$,
although in this model it remains $\lsim30\%$. This result agrees well with
that in Ref. \BT\ for the relevant case ($m_{\tilde q}=2m_{\tilde g}$) in that
paper. When the spoiler mode is open ($m_{\chi^0_2}>m_{\chi^0_1}+h$ or
approximately $m_{\chi^\pm_1}\gsim2m_h$) the dilepton branching ratio vanishes,
although this occurs only for a small portion ($\lsim10\%$) of the allowed
parameter space.

The results for the $\chi^0_2$ dilepton branching ratio for the no-scale
flipped model are shown in Fig. 4 and exhibit quite a bit of structure with
values ranging from zero up to $\approx2/3$. For a given chargino mass, the
various points correspond to different values of $\tan\beta$. For example,
in Fig. 4a for fixed $m_{\chi^\pm_1}\gsim100\GeV$, $\tan\beta$ starts at
the bottom of the pack of curves at $28$ and then decreases in steps of two
until it becomes $4\,(8)$ for $\mu>0\,(\mu<0)$ at the top. (The isolated curve
for $\mu>0$ corresponds to $\tan\beta=2$.) A new spoiler mode opens when
the sneutrino is sufficiently light, but this one and the original spoiler
mode can be overtaken by an `anti-spoiler mode' when the $\tilde l_{L,R}$ are
sufficiently light. In fact, the branching ratio approaches its maximum value
in this case. Clearly the $Z$-exchange diagrams play no dominant role here
since there is no accumulation of points around $6.6\%$.

Finally we compute the number of trilepton events per $100\ipb$ of integrated
luminosity for both models, summed over all possible $e$ and $\mu$
combinations. In the minimal $SU(5)$ supergravity model (see
Fig. 5) we find at least one event for all allowed points in parameter space
(except for when the spoiler mode is open) and as high as $140\,(84)$ events
for $\mu>0\,(\mu<0)$ for low chargino masses. The actual fraction of points
which could be probed at Fermilab depends on the ultimate integrated luminosity
achieved and on the experimental efficiencies for the detection of these
signals. Statistically speaking, only points in parameter space for which
three or more events are predicted could be experimentally verified or
excluded. An across-the-board $30\%$ efficiency cut appears reasonable, with
lower (higher) efficiencies expected for lighter (heavier) chargino masses
\JTW. This situation will probe points in the parameter space with 10 or more
predicted events, \ie,  about half of the allowed parameter space in this
model. An idealized situation would occur with $200\ipb$ (\eg, combining the
data from both detectors) and a $60\%$ efficiency for the heavier chargino
masses. In this case, values down to $2.5$ in Fig. 5 could be probed which
constitutes a large portion of the allowed parameter space and chargino masses
up to $\approx90-95\GeV$. Bearing in mind the mass correlations in this model,
this probe would explore indirectly gluino masses as high as
$320\,(460)\GeV$ for $\mu>0\,(\mu<0)$.

The results for the no-scale flipped model are shown in Fig. 6 (excluding
points for which the spoiler modes are open). Generally we see a
wide ranging number of events for $m_{\chi^\pm_1}\lsim100\GeV$ and a settling
down for heavier chargino masses. The number of events for light charginos
can be quite large, as high as 420 for $m_t=160\GeV$ and $\mu<0$. In the
conservative scenario discussed above one could probe as high
as $m_{\chi^\pm_1}\approx140-160\GeV$ depending on $m_t$, although a large
unexplored region with chargino masses all the way down to the LEP limit  will
remain. In the idealized situation, the unexplored regions will diminish
considerably and the reach could be extended up to
$m_{\chi^\pm_1}\approx200\GeV$. Note that in this model
the range of Fermilab for chargino masses could double that of LEPII. The
indirect reach for $m_{\tilde g}$ due to the mass correlations in this model
could be as high as $m_{\tilde g}\approx750-850\GeV$.

Before we conclude let us also present the number of trilepton events for the
strict case of the no-scale flipped $SU(5)$ supergravity model. As discussed in
Sec. 2.2, this allows us to determine $\tan\beta$ and the sign of $\mu$. The
values for the cross section and branching ratios are more precise here, since
they only depend on $m_t$ for fixed $m_{\chi^\pm_1}$, but they still fall
within the limits found above. In Fig. 7 we show the results. The same general
remarks as for the regular no-scale model apply here, although this time one
could easily {\it exclude} ranges of $m_{\chi^\pm_1}$, since the spoiler modes
are only open for $m_{\chi^\pm_1}\lsim75\GeV$. For the conservative
scenario, it should be possible to fully explore $75\GeV\lsim
m_{\chi^\pm_1}\lsim150\GeV$, whereas in the idealized situation the upper end
could be pushed up to $200\GeV$.

Note that any excluded range of chargino masses or more generally any excluded
portion of the parameter space, implies constraints on the sparticle masses.
In particular, the chargino mass ranges discussed above apply {\it also} to the
second-to-lightest neutralino masses (see Eqns. \III{a} and \IIIf{a}). This
correspondence is quite {\it accurate} for the no-scale flipped model, and
more qualitative for the minimal $SU(5)$ model.

\newsec{Conclusions}
We have studied the signal for the best prospect to detect supersymmetry
at Fermilab, short of actually observing the gluino and squarks. The
neutralino-chargino sector in the class of unified, supersymmetric models
we have explored has the potential of probing a very large portion of the
whole parameter space of these models, without actually seeing the
traditionally sought for strongly interacting supersymmetric particles.
We have considered the minimal $SU(5)$ supergravity model and a recently
proposed no-scale flipped $SU(5)$ supergravity model. These are  well motivated
models which can be taken as typical examples of non-string--like and
string--like models respectively. Results in the class of models we consider,
are expected to differ from those obtained in the ``MSSM" since there: (i) the
large dimension of the parameter space does not allow its complete exploration,
(ii) important theoretical constraints which exclude large regions of parameter
space are not taken into account, and (iii) the interdependence of the various
sparticle and Higgs masses is absent. Our computations of the branching ratios
in the no-scale flipped model provide a clear example of this situation.

We have considered two experimental scenaria which could be realized in due
time and have shown that a large fraction of the parameter space for both
models could be probed. For the minimal $SU(5)$ supergravity model it should
be possible to explore up to $m_{\chi^\pm_1}\approx90-95\GeV$ (and indirectly
$m_{\tilde g}$ as high as $460\GeV$), whereas
in the no-scale flipped model it may be possible to reach up to
$m_{\chi^\pm_1}\approx200\GeV$ (and indirectly $m_{\tilde g}$ as high as
$850\GeV$) and therefore even double the reach of LEPII for chargino and
neutralino masses.

\bigskip
\bigskip
\bigskip
\bigskip
\noindent{\it Acknowledgments}: We would like to thank J. White for very
helpful discussions. This work has been supported in part by DOE
grant DE-FG05-91-ER-40633. The work of J.L. has been supported by an SSC
Fellowship. The work of  D.V.N. has been supported in part by a grant from
Conoco Inc. We would like to thank the HARC Supercomputer Center for the use of
their NEC SX-3 supercomputer and the Texas A\&M Supercomputer Center for the
use of their CRAY-YMP supercomputer.
\vfill\eject
%\listrefsd
\listrefs
\vfill\eject

\noindent{\bf Table I}: Comparison of the most important features describing
the minimal $SU(5)$ supergravity model and the no-scale flipped $SU(5)$
supergravity model.
\bigskip
\input tables
\thicksize=1.0pt
\leftjust
\parasize=2.6in
\begintable
Minimal $SU(5)$ supergravity model\| No-scale flipped $SU(5)$ supergravity
model\crthick
\para{Not easily string-derivable, no known\nextline examples}\|
\para{Easily string-derivable, several known\nextline examples}\nr
\para{Symmetry breaking to Standard Model due to vev of \r{24} and
independent of supersymmetry breaking}\|
\para{Symmetry breaking to Standard Model due to vevs of \r{10},\rb{10}
and tied to onset of supersymmetry breaking}\nr
\para{No simple mechanism for doublet-triplet splitting}\|
\para{Natural doublet-triplet splitting mechanism}\nr
\para{No-scale supergravity excluded}\|{No-scale supergravity by
construction}\nr
\para{$m_{\tilde q},m_{\tilde g}<1\TeV$ by ad-hoc choice:\nextline
naturalness}\|
\para{$m_{\tilde q},m_{\tilde g}<1\TeV$ by no-scale mechanism}\nr
\para{Parameters 5: $m_{1/2},m_0,A,\tan\beta,m_t$}\|
\para{Parameters 3: $m_{1/2},\tan\beta,m_t$}\nr
\para{Proton decay: $d=5$ large, strong\nextline constraints needed}\|
\para{Proton decay: $d=5$ very small}\nr
\para{Dark matter: $\Omega_\chi h^2_0\gg1$ for most of the parameter space,
strong constraints needed}\|
\para{Dark matter: $\Omega_\chi h^2_0\lsim0.25$, ok with cosmology and big
enough for dark matter problem}\nr
\par{$1\lsim\tan\beta\lsim3.5$, $m_t<180\GeV$, $\xi_0\gsim6$}\|
\par{$2\lsim\tan\beta\lsim32$, $m_t<190\GeV$, $\xi_0=0$}\nr
\par{$m_{\tilde g}\lsim500\GeV$}\|
\par{$m_{\tilde g}\lsim1\TeV$,
$m_{\tilde q}\approx m_{\tilde g}$}\nr
\par{$m_{\tilde q}>m_{\tilde l}>2m_{\tilde g}$}\|\par{$m_{\tilde l_L}\approx
m_{\tilde\nu}\approx0.3m_{\tilde g}\lsim300\GeV$}\nr
\par{}\|\par{$m_{\tilde l_R}\approx0.18 m_{\tilde g}\lsim200\GeV$}\nr
\par{$2m_{\chi^0_1}\sim m_{\chi^0_2}\sim m_{\chi^\pm_1}\sim
0.3 m_{\tilde g}\lsim150\GeV$}\|
\par{$2m_{\chi^0_1}\sim m_{\chi^0_2}\approx m_{\chi^\pm_1}\sim
0.3 m_{\tilde g}\lsim285\GeV$}\nr
\par{$m_{\chi^0_3}\sim m_{\chi^0_4}\sim m_{\chi^\pm_2}\sim\vert\mu\vert$}\|
\par{$m_{\chi^0_3}\sim m_{\chi^0_4}\sim m_{\chi^\pm_2}\sim\vert\mu\vert$}\nr
\par{$m_h\lsim100\GeV$}\|\par{$m_h\lsim135\GeV$}\cr
\par{}\|\par{Strict no-scale: $\tan\beta=\tan\beta(m_{\tilde g},m_t)$}\nr
\par{No analog}\|{$m_t\lsim135\GeV\Rightarrow\mu>0,m_h\lsim100\GeV$}\nr
\par{}\|{$m_t\gsim140\GeV\Rightarrow\mu<0,m_h\gsim100\GeV$}

\endtable

%\listfigsd
\listfigs
\bye